\documentstyle[11pt,aaspp4,tighten]{article}

\begin{document}

\title{Three-Dimensional Evolution of the Parker Instability
under a Uniform Gravity}
\author{Jongsoo Kim\altaffilmark{1,2}, 
        S. S. Hong\altaffilmark{2},
        Dongsu Ryu\altaffilmark{3},
        and T. W. Jones\altaffilmark{4}}

\altaffiltext{1}
{Korea Astronomy Observatory, San 36-1, Hwaam-Dong, Yusong-Ku,
Taejon 305-348, Korea: \\jskim@hanul.issa.re.kr}
\altaffiltext{2}
{Department of Astronomy, Seoul National University, Seoul 151-742, 
Korea: \\sshong@astroism.snu.ac.kr}
\altaffiltext{3}
{Department of Astronomy and Space Science, Chungnam National 
University, Taejon 305-764, Korea: \\ryu@canopus.chungnam.ac.kr}
\altaffiltext{4}
{Department of Astronomy, University of Minnesota, Minneapolis, MN 
55455: \\twj@msi.umn.edu} 

\begin{abstract}
 
Using an isothermal MHD code, we have performed three-dimensional, 
high-resolution simulations of the Parker instability.  The initial 
equilibrium system is composed of exponentially-decreasing isothermal gas 
and magnetic field (along the azimuthal direction) under a uniform gravity. 
The evolution of the instability can be 
divided into three phases: {\it linear}, {\it nonlinear}, and {\it relaxed}.
During the linear phase, the perturbations grow exponentially with a
preferred scale along the azimuthal direction but with smallest possible scale
along the radial direction, as predicted from linear analyses.  During the 
nonlinear phase, the growth of the instability is saturated and flow motion 
becomes {\it chaotic}.  Magnetic reconnection occurs, which allows gas to 
cross field lines.  This, in turn, results in the redistribution of gas and 
magnetic field.  The system approaches {\it a new equilibrium} in the 
relaxed phase, which is different from the one seen in two-dimensional works.
The structures formed during the evolution are sheet-like or filamentary, 
whose shortest dimension is radial.  Their maximum density enhancement factor 
relative to the initial value is less than 2.  Since the radial dimension is 
too small and the density enhancement is too low, it is difficult to regard
the Parker instability {\it alone} as a viable mechanism for the formation 
of giant molecular clouds.   

\end{abstract}


\clearpage

\section{INTRODUCTION} 

Under a uniform gravity, a vertically stratified system of interstellar medium 
(ISM) composed of gas, magnetic field and cosmic-rays is unstable against 
perturbations (Parker 1966, 1967).   This nature has been called the 
{\it Parker instability} in the astronomical literature.  The instability
is broken down into two distinct modes: {\it undular} and {\it interchange} 
(Hughes \& Cattaneo 1987).  The former, whose wave vectors lie in the plane 
defined by the direction of gravitational field (the vertical direction) 
and the direction of unperturbed magnetic field (the azimuthal direction), 
undulates the azimuthal field lines and induces the gas to slide down along 
the field lines into ``magnetic valleys''.  The critical adiabatic index for 
the undular mode without taking into account cosmic-rays is 
$\gamma_{\rm cr,u} = (1+\alpha)^2/(1+1.5\alpha)$, where $\alpha$ is the 
ratio of magnetic to gas pressure.
The undular mode has a preferred wavelength 
along the azimuthal direction.  The latter interchange mode, whose wave 
vectors lie in the plane perpendicular to the direction of unperturbed 
magnetic field, leaves the direction of the magnetic field unchanged
but alters its strength.  The critical adiabatic index of the interchange 
mode is  $\gamma_{\rm cr,i}=1-\alpha$.  This mode has a maximum growth 
rate at infinite wavenumber, indicating it is a Rayleigh-Taylor type.   
If the wave vectors are allowed to have all three components,
then there exists the {\it mixed mode} of the undular and interchange modes
(Matsumoto et al. 1993). The critical 
adiabatic index of the mixed mode is $\gamma_{\rm cr,m}=1+\alpha$.

If three-dimensional (3D) perturbations are applied to a system with the
{\it effective} adiabatic index $\gamma_{\rm cr,i}<\gamma<\gamma_{\rm cr,u}$,
which is typical of the ISM, the system is unstable 
to the undular mode but stable to the interchange one.  The mixed mode, 
however, has maximum growth rate at vanishing wavelength along the
radial direction (Parker 1967), showing the characteristic of the 
interchange mode.  This implies that the interchange mode as well as 
the undular mode are involved in the 3D development of 
the Parker instability. It is our understanding that the ``magnetic arches'', 
which are formed by the undular mode, are less dense than their
surroundings, and are now susceptible to the interchange mode.

It has been considered that the Parker instability is one of the plausible
mechanisms for the formation of giant molecular clouds (GMCs) 
(Mouschovias et al. 1974; Blitz \& Shu 1980).  
As the formation mechanism of GMCs, the Parker instability, however, 
has two problems. One is that, as we have mentioned in the last paragraph,
its linear growth rate is maximum at infinite wavenumber along the radial
direction of our Galaxy.  This means that chaotic structures
rather than ordered large-scale ones are expected to form.
The other is that, according to two-dimensional numerical simulations
of the Parker instability (Basu et al. 1997), the density enhancement
is at most a factor of 2 or so, which is too small.

In this paper, we report the 3D simulations of the
Parker instability under a uniform gravity, which are intended to
study the mixed mode.  Our work is a 3D extension of 
that of the two-dimensional undular mode by Basu et al. (1996, 1997).
The simulations enable us to answer the following questions on
the 3D development of the Parker instability:
i) What are the structures formed in the nonlinear stage?
ii) What is the density enhancement factor?
Based on the answers, we address the role of the Parker instability
in the formation of GMCs.  In addition, the simulations show that there 
is a final equilibrium state which is different from that obtained in 
the two-dimensional study (Mouschovias 1974; Basu et al. 1996, 1997).
Other numerical works on the Parker instability
include two-dimensional simulations of the undular mode by
Matsumoto et al. (1988, 1990) and 3D simulations
of the mixed mode by Matsumoto \& Shibata (1992),
both of which employed a point-mass-dominated gravity.
The point-mass-dominated gravity can mimic the linearly increasing part
of the Galactic gravity at solar neighborhood when the $\epsilon$ parameter
(the ratio of gravitational energy to thermal plus magnetic energy)
becomes as large as about 1000.  Their simulations under the 
gravity correctly demonstrated the nonlinear 
development of the undular and mixed modes, although their numerical 
resolution was lower than that of Basu et al. and ours.   
However, application of their results to the Galactic disk was hampered, 
because their parameter ($\epsilon=6$) is too small to model the Galactic
gravity.

In \S 2, the initial setup and the numerical code are briefly described.
The results of the simulations and discussions are presented in \S 3.

\section{NUMERICAL SETUP AND CODE}

To describe the local behavior of the Parker instability, we introduce 
Cartesian coordinates $(x,y,z)$ whose directions are radial, azimuthal 
and vertical, respectively.  Under a uniform external gravity, 
$-g\hat{z}$, the initial magnetohydrostatic equilibrium state of a system 
composed of isothermal gas with sound speed, $a$, and uni-directional 
magnetic field, $B_0 = \hat{y}B_0(z)$, is given by  
\begin{equation}\label{istate}
\frac{\rho_0(z)}{\rho_0(0)} =
\frac{B_0^2(z)}{B_0^2(0)} =
\exp(-z/H),
\end{equation}
where the vertical scale height of the gas is defined as $H = (1+\alpha)a^2/g$.
As Parker (1966) did, we assume $\alpha (\equiv B_0^2/[8\pi \rho_0 a^2])$
to be uniform initially.  We study the case with the initial $\alpha=1$ 
and the adiabatic index $\gamma=1$.

The computational box has $0 \leq x \leq 12H$, $0 \leq y \leq 12H$
and $0 \leq z \leq 12H$, where $12H$ is the azimuthal wavelength of maximum 
linear growth (Parker 1966).  The periodic boundary condition is enforced 
at $x=0$, $x=12H$, $y=0$ and $y=12H$, while the reflection boundary 
condition at $z=0$ and $z=12H$.  To initiate the instability in the 
initial equilibrium state, we add random velocity perturbations.
The standard deviation of each velocity component is set to
be $10^{-4}a$.  Since the 3D mixed mode has maximum linear 
growth at infinite wavenumber along the radial direction, high resolution 
simulations are required in order to resolve resulting small scale structures.
We have performed simulations using $64^3$, $128^3$ and $256^3$ cells,
although we report mostly the highest resolution one with $256^3$ cells.
Physical quantities of length, speed, density and magnetic field are given 
in units of the density scale height, $H$, the isothermal sound speed, $a$, 
the initial density at $z=0$, $\rho_0(0)$, and the initial field strength at 
$z=0$, $B_0(0)$, respectively.
For the typical values of $a$=6.4~km~s$^{-1}$ and $H$= 160~pc (Falgarone
\& Lequeux 1973), the resulting
time unit, $H/a$, becomes $2.5\times10^7$~years.
The simulations have been made with a 3D isothermal MHD code
based on the explicit, finite-difference TVD (Total Variation Diminishing)
scheme (Kim et al. 1998).  

\section{RESULTS AND DISCUSSIONS}

Figure 1 shows the time evolution of the logarithmic values of
the root-mean-square (rms) velocity components 
in the simulations with $128^3$ and $256^3$ cells.
Two solid lines are the maximum linear growth rates of the mixed 
mode, $\Omega_{\rm max, m} = 0.41$, and the undular mode, 
$\Omega_{\rm max, u} = 0.34$ (Parker 1966, 1967).
The evolution patterns in the two panels agree well, indicating
the simulations produce converged results at least for the global
quantities such as $\langle v^2\rangle^{1/2}$.
The evolution is divided into three phases: 
linear ($ 0 \leq t\stackrel{<}{_\sim}35$), 
nonlinear ($35\stackrel{<}{_\sim}t\stackrel{<}{_\sim}50$) 
and relaxed ($t\stackrel{>}{_\sim}50$).

During the initial transient stage of the linear phase 
($0 \leq t \stackrel{<}{_\sim} 25$), random velocity perturbations adjust
themselves by generating a one-dimensional magnetosonic wave, which
moves vertically.  The peaks in $\langle v_z^2\rangle^{1/2}$ around 
$t\simeq7$ and $21$ occur when the
wave reaches the upper boundary where background density is lowest.
After the transient stage, the instability grows exponentially
($25\stackrel{<}{_\sim}t\stackrel{<}{_\sim}35$).   
Kim et al. (1998) showed that in a two-dimensional test simulation
the Parker instability grows at the maximum linear
growth rate of the undular mode, $\Omega_{\rm max, u} = 0.34$.
In Figure 1, however, the Parker instability in 3D simulations grows at a rate
that is smaller than the maximum linear growth rate of the mixed mode,
although slightly larger than that of the undular mode.
This is because the maximum growth of the mixed mode occurs at
infinite wavenumber along the radial direction, while numerical
simulations with finite resolution can not resolve such small structures.

The growth of the instability is slowed down after $t\stackrel{>}{_\sim}35$
and saturated in the nonlinear phase due to magnetic tension.
The 3D images in Figure 2 depict density structure,
velocity vectors and magnetic field lines at a nonlinear epoch,
$t=40$, (Figure 2a) as well as at a relaxed epoch, $t=60$ (Figure 2b).
The two-dimensional images in Figure 3 depict density structure and
velocity vectors in a sliced $yz$ plane, $xz$ plane, and $xy$ plane at $t=40$,
when the instability is fully developed.  Figure 2a shows magnetic 
arches and valleys created along the magnetic field lines.
Their azimuthal wavelength is about 12, that of the undular mode.
The structures in the sliced $yz$ plane of Figure 3 have similarities
with those in two-dimensional simulations of the undular mode
(Basu et al. 1996; Kim et al. 1998).  These are the manifestations of 
the undular mode in the 3D evolution of the Parker instability.
On the other hand, the iso-density surface in Figure 2a is corrugated
along the $x$-direction.  Also, in the sliced $xz$ and $xy$ planes, 
the density and velocity changes rapidly along the direction.
These are the manifestations of the interchange mode.  The structure has 
maximum power at the $x$-wavenumbers corresponding to $1/16$ to $1/32$ 
of the box size, or $8$ to $16$ cells.  This is the size of the smallest 
structures the code can resolve cleanly (Ryu et al. 1995; Kim et al. 1998).
Structures smaller than that tend to be dissipated by numerical diffusion.
Simulations with higher resolutions would have produced structures
with larger $x$-wavenumbers.  Comparison between the results of the high 
resolution simulation using $256^3$ cells and that of the medium resolution 
simulation using $128^3$ cells has confirmed the above statement.
Hence, we can state that due to the manifestation of the interchange mode, 
the flow motion in the 3D simulations should be characterized as chaotic.
This is different from what was seen in two-dimensional works
(Mouschovias 1974; Basu et al. 1996, 1997).

The chaotic flow motion in the nonlinear phase augments
magnetic reconnection, especially in the magnetic valley regions
(see Jones et al. 1997 for further discussion on reconnection in simulations
of this kind).
As it accumulates, gas in the valleys is compressed and magnetic
field lines are pushed down by the weight of the compressed gas,
generating strongly curved field lines.  Subsequently, the curved field 
lines get reconnected, facilitated by the chaotic flow motion.
The reconnection allows the gas to drop down off the field lines, and
at the same time the reconnected field lines to float upwards.
In this way, gas aggregates along the midplane while
magnetic field is evacuated from it.  The chaotic flow motion induces 
reconnection in other parts of the computational box, but there it 
contributes less in the redistribution of gas and magnetic field.
During the relaxed phase, the dense gas with higher pressure than its
surroundings spreads along the magnetic field lines, and the redistributed
gas and magnetic field flatten themselves horizontally.
A snapshot of the relaxed phase in Figure 2b shows that by $t=60$ the
corrugation has been smeared out, the velocity field has been almost
aligned with the magnetic field lines, and the magnetic field lines
have become more or less straight.
However, the iso-density surface, which coincides initially with the
$z=3$ plane, has moved downwards to $z\simeq2$.
This is an indication that the relaxed system is significantly different
from the initial one.

To show quantitatively how much gas is segregated from magnetic field,
we plot $\bar{\alpha}$, 
which is defined as
\begin{equation}\label{alpha}
\bar{\alpha}(z;t) = 
\int_{0}^{12H}\!\!\int_{0}^{12H} 
\frac{B^2(x,y,z;t)}{8\pi a^2 \rho(x,y,z;t)}dxdy \; / \; 
\int_{0}^{12H}\!\!\int_{0}^{12H}dxdy,
\end{equation}
with respect to $z$ at several epochs in Figure 4.  Initially $\bar{\alpha}$  
is equal to 1.  As time goes by, $\bar{\alpha}$ decreases near the midplane, 
because gas aggregates there while magnetic field is evacuated as
stated in the above paragraph.  By $t=60$, $\bar{\alpha}$ decreases by a 
factor of 10 in the midplane.  It increases  more than by a factor of 10 
around the upper boundary.  At $t=60$, in the range 
$0 \leq z \stackrel{<}{_\sim} 4$ magnetic field is more or less uniform and 
the external gravity is now supported almost totally by gas pressure.  But 
for $z \stackrel{>}{_\sim} 4$, $\sim90\%$ of the external gravity is supported 
by magnetic pressure and the rest by gas pressure.

So, if the Parker instability is allowed to develop into the relaxed phase
without being interrupted by other events, such as explosive energy injections
by OB associations in the Galactic plane, the initial unstable Parker
system should evolve into a new stable equilibrium state where the external
gravity is mostly supported by the gas pressure alone.
In other words, one role that the Parker instability can play in the
Galactic plane is to reduce the scale height of interstellar clouds while
increasing that of interstellar magnetic field.

Energetics in our simulations has been checked through energy plots
(not shown in this paper) which are similar to those in Matsumoto et al.
(1988, 1990).  The total energy of the system (see Mouschovias 1974 for
definition) stays constant during the linear phase, but decreases
significantly during the nonlinear phase.  This is because
the total energy is no longer a conserved quantity in an isothermal
flow, when shock waves are generated in the system (Matsumoto et al. 1990).

How long does it take the Parker instability under the uniform gravity to grow?
In estimating the elapsed time, the initial transient stage should not
be taken into account, since its duration depends on initial perturbations 
and can be reduced if we start simulations with an initial setup prescribed 
by the linear analysis.  The density at magnetic valleys starts to increase 
significantly only after the initial transient stage. 
Hence, the duration of the linear phase except the initial transient
stage may be regarded as the growth time of the instability in our simulations.
It amounts to $\sim2.5\times10^8$ years. 
What is the density enhancement resulting from the 3D 
Parker instability?
At the end of the linear phase, $t\simeq35$, the density enhancement
factor reaches the maximum value. Later, it slowly decreases.
On the scale of GMCs, $\sim 60$ pc, which corresponds to
8 cells in the high resolution simulation, the maximum density
enhancement along the midplane is $\stackrel{<}{_\sim} 2$ at $t=35$.

The Parker instability has been considered as one of the plausible
mechanisms for the formation of GMCs as described in Introduction.
Based on our 3D simulations of the Parker instability
under the uniform gravity, we argue the following: The time scale, 
$\sim2.5\times10^8$ years, required for the development of the instability is 
somewhat larger than the lifetime ($\sim3\times10^7$ years) of GMCs 
(Blitz \& Shu 1980).  The time scale can, however, be reduced, if we 
replace the uniform gravity by a realistic gravity (Kim \& Hong 1998).  
Therefore, as far as time scale is concerned, the Parker instability can be 
regarded a viable formation mechanism of GMCs.  However, the structures 
formed in the 3D simulations are sheet-like or 
filamentary, and the maximum density enhancement factor in the scale of
GMCs is only $\sim2$.  
Hence, we conclude that {\it it is difficult to regard the Parker 
instability alone as the formation mechanism  of GMCs.} 

\acknowledgments

We are grateful to Drs. B.-C. Koo and Y. Lee for comments on the manuscript. 
Computations in the present work were carried out by using CRAY T3Es at
SERI in Korea and at the University of Minnesota Supercomputing
Institute in the US. 
The work by JK was supported in part by the Ministry of Science and
Technology through Korea Astronomy Observatory grant 97-5400-000.
The work by SSH was supported in part by the Korea Research Foundation
made in the year of 1997.
The work by DR was supported in part by KOSEF through grant
981-0203-011-2.
The work by TWJ was supported in part by NSF grants AST-9619438 and
INT-9511654 and by the University of Minnesota Supercomputing Institute.

\clearpage

\centerline{\bf FIGURES}

\figurenum{1}
\figcaption{
Time evolution of the root-mean-square values of the radial velocity,
$<v_x^2>^{1/2}$, the azimuthal velocity, $<v_y^2>^{1/2}$, and the vertical
velocity, $<v_z^2>^{1/2}$.  Natural log is used along the ordinate.
Left panel is from the medium resolution simulation with $128^3$ cells,
and right panel is from the high resolution simulation with $256^3$ cells.
Two solid lines represent the maximum linear growth rates, 0.34 and 0.41,
of the undular mode and the mixed mode, respectively.  
}

\figurenum{2}
\figcaption{
Perspective views of density structure, velocity vectors and magnetic field 
lines at two time epochs, (a) $t=40$, and (b) $t=60$, from the high resolution 
simulation with $256^3$ cells.  Box is oriented in such a way that the 
radial ($x-$), the azimuthal ($y-$), and  the vertical ($z-$) directions 
are from left to right, from near to far, and from bottom to top, respectively.
Translucent images of density in the three planar slices, $xy$ plane at 
$z=3$, $yz$ plane at $x=6$, and $zx$ plane at $y=6$, are overlapped.  
Colors are mapped from red to blue as density decreases.
Gray surface with an iso-density of $\rho_0(z=3)$ is included.
Velocity field vectors in the $xy$ plane at $z=3$ are represented by
red arrows.  Magnetic field lines, whose starting points lie along
the line of $z=6$ and $y=0$, are represented by blue ribbons
}

\figurenum{3}
\figcaption{
Images of density structure and velocity vectors on the three planes, 
$x=6$ (left), $y=6$ (center), and $z=3$ (right) at $t=40$ from the high 
resolution simulation with $256^3$ cells.  The three planes are same as the 
sliced planes in Figure 2.  Colors are mapped from red to blue as density 
decreases.  The unit of the velocity vectors is shown at the right lower 
corner of the image.   
}

\figurenum{4}
\figcaption{
The ratio of magnetic to gas pressure $\bar{\alpha}$ averaged over
$xy$ plane (see eq.~[\ref{alpha}]) at several time epochs 
from the high resolution simulation with $256^3$ cells.  
Common log is used along the ordinate.
}

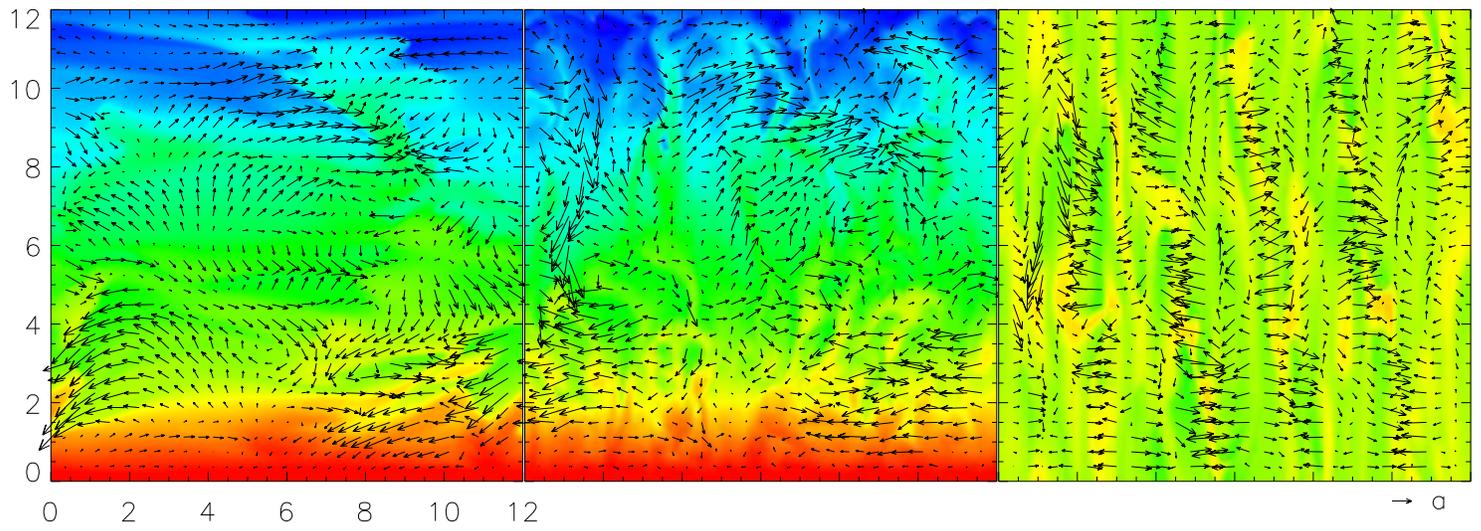
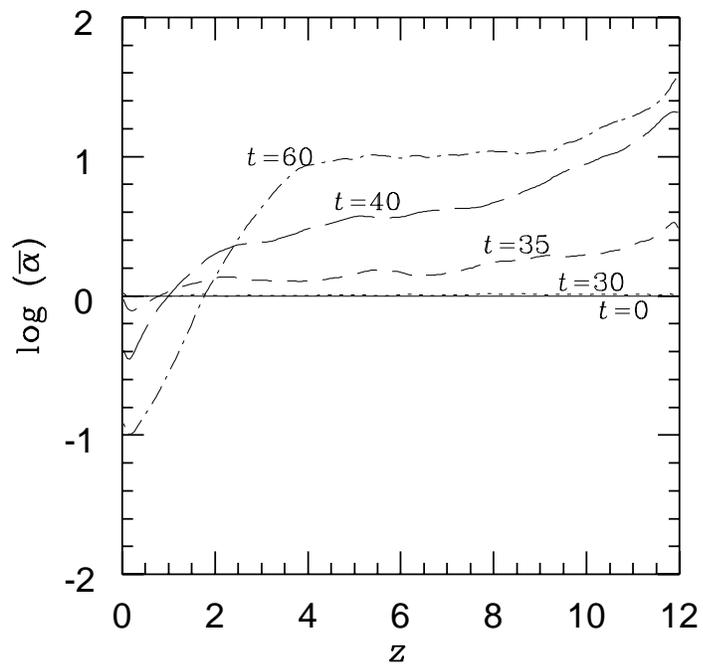
\end{document}